\documentclass[aps,prx,longbibliography,showpacs,amsmath,amssymb,amsfonts,twocolumn]{revtex4-1}

\usepackage{graphicx}
\usepackage{dcolumn}
\usepackage{bm}
\usepackage{epsf}
\usepackage{verbatim}

\newcommand{\tr}[1]{\mathrm{tr}\left\{#1\right\}}

\newcommand{\la}{\left\langle}
\newcommand{\ra}{\right\rangle}
\newcommand{\pd}{\partial}

\newcommand{\td}{\mathrm{d}}

\newcommand{\e}[1]{\exp{\left(#1\right)}}

\newcommand{\co}[1]{\cos{\left(#1\right)}}
\newcommand{\si}[1]{\sin{\left(#1\right)}}

\newcommand{\bla}{bla\\bla\\bla\\bla\\bla}

\newcommand{\PRA}{Phys. Rev. A}

\newcommand{\PRE}{Phys. Rev. E}
\newcommand{\PRL}{Phys. Rev. Lett.}
\newcommand{\EPL}{EPL (Europhys. Lett.)}
\newcommand{\RMP}{Rev. Mod. Phys.}
\newcommand{\NJP}{New. J. Phys.}

\newcommand{\mb}[1]{\mbox{\boldmath$#1$}}
\newcommand{\mc}[1]{\mathcal{#1}}

\newcommand{\mf}[1]{\mathfrak{#1}}
\newcommand{\mrm}[1]{\mathrm{#1}}

\def\dbar{{\mathchar'26\mkern-12mu {\rm d}}}

\begin{document}

\title{Information processing and the second law of thermodynamics: an inclusive, Hamiltonian approach} 

\author{Sebastian Deffner}
\email{sebastian.deffner@gmail.com}
\author{Christopher Jarzynski}
\email{cjarzyns@umd.edu}
\affiliation{Department of Chemistry and Biochemistry and Institute for Physical Science and Technology, University of Maryland, 
College Park, Maryland 20742, USA}
\date{\today}

\begin{abstract}
We obtain generalizations of the Kelvin-Planck, Clausius, and Carnot statements of the second law of thermodynamics, for situations involving information processing.  To this end, we consider an {\it information reservoir} (representing, e.g.\ a memory device) alongside the heat and work reservoirs that appear in traditional thermodynamic analyses.  We derive our results within an inclusive framework in which all participating elements -- the system or device of interest, together with the heat, work and information reservoirs -- are modeled explicitly by a time-independent, classical Hamiltonian.  We place particular emphasis on the limits and assumptions under which cyclic motion of the device of interest emerges from its interactions with work, heat, and information reservoirs.
\end{abstract}


\maketitle

Three classic expressions of the second law of thermodynamics are formulated in terms of cyclic processes. The Kelvin-Planck statement asserts that \cite{thomson_82,planck_54}
\begin{quote}
\textit{no process is possible whose sole result is the extraction of energy from a heat bath, and the conversion of all that energy into work.}
\end{quote}
The Clausius statement reads \cite{clausius_64},
\begin{quote}
\textit{no process is possible whose sole result is the transfer of heat from a body of lower temperature to a body of higher temperature.}
\end{quote}
Finally, the Carnot statement declares that \cite{carnot_24}
\begin{quote}
\textit{no engine operating between two heat reservoirs can be more efficient than a Carnot engine operating between those same reservoirs.}
\end{quote}
These formulations refer to processes involving the exchange of energy among idealized subsystems:
one or more {\it heat reservoirs}; a {\it work source} -- for example, a mass that can be raised or lowered against gravity; and
a {\it device} that operates in cycles, and effects the transfer of energy among the other subsystems.
All three statements follow from simple entropy balance analyses, and offer useful, logically transparent reference points as one navigates the application of the laws of thermodynamics to real systems.

This paper concerns extensions of these classic statements to situations involving information processing. In addition to the above-mentioned elements, we will consider an {\it information reservoir} -- a system that exchanges information but not energy with the device. As we will show, the Kelvin-Planck, Clausius, and Carnot statements are each generalized in a natural way in the presence of such a reservoir. Although these generalized statements can be derived {\it ad hoc}, simply by including the Shannon entropy of the information reservoir in the entropy balance analysis, our aim is to obtain these results directly from microscopic, Hamiltonian dynamics, highlighting the assumptions and approximations that are made along common idealizations.

Among the various connections that exist between information theory and thermodynamics, two are relevant in the present context.
The first involves the relationship between the thermodynamic entropy defined via the Clausius relation, $\int \dbar Q/T = \Delta S$, and the Shannon entropy of information theory~\cite{sha48},
\begin{equation}
\label{eq:Shannon_entropy}
\mc{H}=-\tr{\rho\ln{\rho}} \equiv - \int \rho \ln \rho \, ,
\end{equation}
where $\int$ denotes an integral over phase space. Since these definitions coincide for a system in canonical equilibrium with a heat reservoir~\cite{hill56}, it is highly tempting to use Eq.~\eqref{eq:Shannon_entropy} to define the entropy of a {\it nonequilibrium} state.
Indeed, if a system in contact with one or more thermal reservoirs evolves from an initial statistical state $\rho_{\rm i}$ to a final state $\rho_{\rm f}$
-- neither of which is assumed to correspond to thermal equilibrium -- then the Clausius-like inequality
\begin{equation}
\label{e01}
\int_{\rm i}^{\rm f} \frac{\dbar Q}{T} \le -\tr{\rho_{\rm f}\ln{\rho_{\rm f}}} + \tr{\rho_{\rm i}\ln{\rho_{\rm i}}} \equiv \Delta\mc{H}
\end{equation}
can be established from microscopic principles, as shown in Ref.~\cite{jar99} under assumptions similar to those we will make in the present paper;
see also Refs.~\cite{pie00,qia02,mae03,sei05,and08,and09,vai09,has10,esp10a,tak10,esp11,sag12} for related results and alternative derivations.
On the other hand, for an isolated classical system, the Shannon entropy $\mc{H}$ remains constant with time (by Liouville's theorem), which conflicts with the observation that the entropy of an isolated physical system increases until equilibrium is attained.
Such considerations show that, at the very least, one must be careful when identifying Shannon entropy with thermodynamic entropy, away from thermal equilibrium. 

The second connection involves the question of whether information about molecular-scale motions, gained by external observation, can be used to \textit{subvert} the second law, in the sense suggested by Maxwell's famous thought experiment~\cite{maxwell_1871}.
In an illuminating refinement of the Maxwell demon framework, Le\' o Szil\' ard described a hypothetical scenario in which an intelligent being takes advantage of microscopic observations to manipulate a single-particle gas, so as to extract energy systematically from a reservoir and convert it to work~\cite{szi29}.
Szil\' ard explicitly raised the possibility that this human-like intelligence could be replaced by a purely physical device, in apparent violation of the Kelvin-Planck statement.
By current consensus, the resolution of this paradox resides in Landauer's principle~\cite{lan61}, which assigns a minimal thermodynamic cost to the erasure of the information gathered by the device; see Refs.~\cite{penrose_1970,ben82,leff_2003} for details, Ref.~\cite{ber12} for an experimental treatment, Refs.~\cite{qua06,bie12,man12,sag12,hor12,str13,bar13,man13} for illustrative models, and Refs.~\cite{ear98,ear99,hem10,nor11} for dissenting perspectives.

These topics have gained recent prominence in the context of microscopic {\it feedback control}. Sagawa and Ueda have analyzed the amount of work that can be delivered by the measurement and manipulation of small, fluctuating systems~\cite{sag08,sag09,sag10,sag11,sag12a}. Their predictions have been verified experimentally using trapped colloidal particles~\cite{toy11} and mathematically illustrated for a solvable system with a linear feedback protocol with a Kalman filter \cite{fuj10}. Their analyses are based on an approach considering an integral fluctuation theorem. The corresponding detailed fluctuation theorems were discussed in Ref.~\cite{pon10,hor10,hor11,hor11a}. These results have been extended to systems prepared in initial non-equilibrium stationary states~\cite{abr11,abr12,lah12,def12} and to quantum systems \cite{mor11,ved12,kav12}. Alternative treatments of feedback control, which do not rely on fluctuation theorems, can be found in Refs.~\cite{kim04,kim07,cao09,suz09,jac09} with applications to theoretic models \cite{fra04,fei09,cao09a,vai11} and experimental systems \cite{lop08,bon09}.

In the feedback control paradigm, the microscopic state of the system of interest (or of a measurement device~\cite{gra11}) is observed, and on the basis of those observations a protocol is adapted to manipulate the system. Implicit in this paradigm is an external agent or apparatus -- the demon or \textit{feedback controller}~\cite{sag12} -- who makes these observations and implements the feedback. The results derived within this approach are thus expressed as relationships between thermodynamic quantities such as work, and information-theoretic quantities that measure the quality of the observations.

In this paper we aim at a treatment that does not involve an external agent.
Instead, we consider a self-contained \textit{universe}, a composite system containing the elements mentioned earlier: a device, one or more thermal reservoirs, a work source, and an information reservoir.
This composite system evolves autonomously under Hamilton's equations of motion, and any effective feedback control arises entirely from the interplay of the subsystems.
Within this \textit{inclusive} framework we will obtain inequalities that generalize the Kelvin-Planck, Clausius, and Carnot statements, to processes involving  the exchange of information.

We will begin in Section~\ref{sec:framework} by specifying our theoretical framework and terminology. In Section~\ref{sec:ft} we will obtain obtain formal inequalities, which will then be combined in Section~\ref{sec:2ndlaw} with physical interpretations, to obtain generalized statements of the second law for cyclic process. In order to complete the analysis we will derive a generalized maximum work theorem for non-cyclic process in Section~\ref{sec:maxwork}. Finally, we will conclude in Section~\ref{sec:conclusions}. In obtaining these results we will make a number of assumptions and approximations, reflecting idealizations that commonly arise in analyses of thermodynamic principles, and we will discuss the roles of these assumptions in our treatment.

\section{Thermodynamics within a Hamiltonian framework}
\label{sec:framework}

\begin{figure}
\centering
 \includegraphics[width=0.48\textwidth]{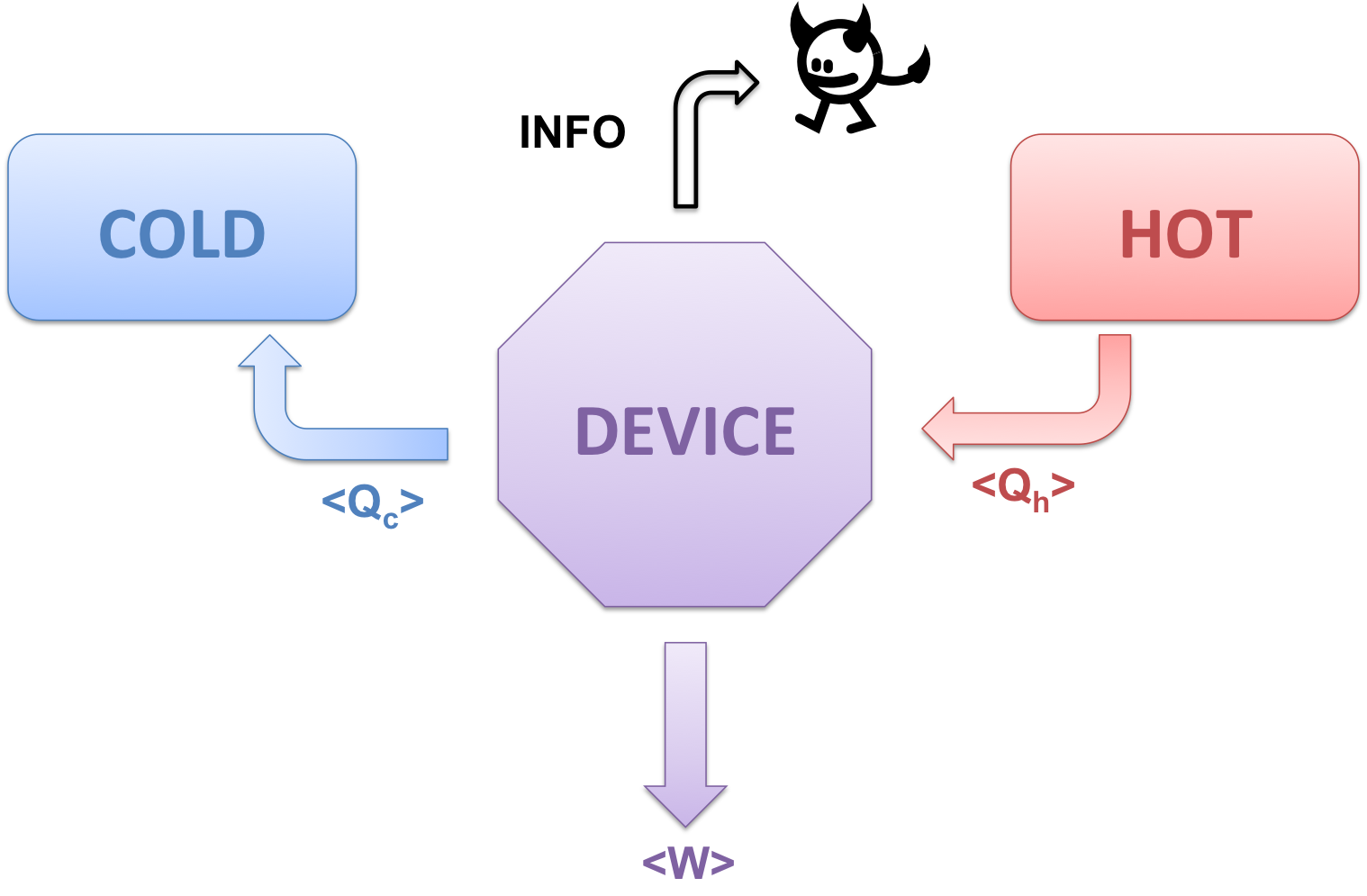}
\caption{\label{thermo} (color online) Thermodynamic set-up: a device exchanges heat with thermal reservoirs and work with a work source. The process is observed by a \textit{Maxwell demon}.}
\end{figure}

In this section we describe our framework, beginning with concepts and terminology. For our purposes, systems are categorized as devices, heat sources, work sources, and information sources. It is important to understand this categorization as an idealization of \textit{real} physical systems with one dominant behavior. 

A {\it heat source} (or {\it sink}) is a system that exchanges energy with other systems, in the form of heat but not work.
Relaxation processes within a heat source are generally assumed to occur rapidly, implying that its temperature remains well defined throughout any process under consideration~\cite{callen_85}.
Moreover, if its heat capacity is sufficiently large, then that temperature remains effectively constant and the heat source can be viewed as a heat {\it reservoir}.
Because we wish to make contact with the usual formulations of the Kelvin-Planck, Clausius,  and Carnot statements, we will use the term heat reservoir in the analysis that follows. In particular, we will see that the assumption of a large heat capacity is crucial for the emergence of cyclic motion of the device.

Analogously, a {\it work source} (or {\it sink}) can exchange energy in the form of work, but not heat, and its internal relaxation processes are again assumed to be rapid~\cite{callen_85}.
As a result, the entropy of the work source remains constant and can be neglected.
The assumption of rapid relaxation will appear implicitly in our treatment:
we will model the work source as a single degree of freedom -- effectively a collective coordinate such as the center of mass of a macroscopic system -- while ignoring its internal degrees of freedom;
this prevents the energy of the work source from being lost to internal dissipation.
If the inertia of a work source is sufficiently large, then its motion is largely unaffected by interactions with other systems, and it can  be viewed as a work {\it reservoir}.
In order to model cyclic processes, we will assume a work source with an effectively infinite inertia; we will return to this point in greater detail shortly.

Heat and work sources are convenient conceptual idealizations, familiar from classic thermodynamic treatments~\cite{fermi_38,planck_54}.
To this list we add an {\it information source}, a system that exchanges information but not energy with other systems.
By this we mean that the information source can exist in a number of physically distinct accessible states with identical free energies.
A useful example is a memory register with $N$ bits, hence $2^N$ energetically degenerate states.
The capacity of the information source is measured by the natural logarithm of the number of accessible states, i.e.\ $N \ln 2$ in the case of the memory register. When this capacity is sufficiently large, the information source becomes an information {\it reservoir}.

Finally, we will consider a {\it device}, or (sub-)system of interest, which interacts with the above-mentioned elements.
These interactions give rise to the exchange of energy with the heat and work sources, and they influence the dynamics of the information source among its degenerate states.
As a result, data relating to the evolution of the device may become encoded in the information source.
It is precisely this possibility that adds a new element to the standard analyses of the Kelvin-Planck, Clausius,  and Carnot statements.

\subsection{First law of thermodynamics}
\label{subsec:firstLaw}

We proceed by formulating the first law of thermodynamics within our classical, Hamiltonian framework.
To begin, we restrict the analysis to include only a device, a work source, and a single heat source (later we will add multiple heat sources and an information source), and we model these elements with a Hamiltonian
\begin{equation}
\begin{split}
\label{e02}
H_\mathrm{TOT}(\gamma)
=H_0(\mb{x},\mb{p};\,X)+h(\mb{x},\mb{p};\,\mb{\xi},\mb{\varphi})\\
+H_\mathrm{work}(X,P)+H_\mathrm{heat}(\mb{\xi},\mb{\varphi})\,,
\end{split}
\end{equation}
with $\gamma = (\mb{x},\mb{p};\,\mb{\xi},\mb{\varphi};X,P)$.
Here $(\mb{x},\mb{p})$ denotes the microstate of the device, and $(\mb{\xi},\mb{\varphi})$ is that of the heat reservoir. The bold letters indicate vectors in the configuration and momentum spaces of these subsystems.
We use $(X,P)$ to specify the microstate of the work reservoir, which we model with a single degree of freedom.
Finally, $\gamma$ denotes a point in the full phase space, describing the combined microstate of all three subsystems.

We view the first term on the right side of Eq.~\eqref{e02}, $H_0$, as the bare Hamiltonian for the device, parametrized by the configuration of the work source, $X$.
The second term, $h$, gives the interaction between the device and the heat source, and the third and fourth terms are the bare Hamiltonians for the work and heat sources.
Defining
\begin{equation}
\label{eq:defHdev}
H_\mathrm{dev}(\mb{x},\mb{p};\mb{\xi},\mb{\varphi};\,X) \equiv H_0(\mb{x},\mb{p};\,X)+h(\mb{x},\mb{p};\,\mb{\xi},\mb{\varphi}) \, ,
\end{equation}
we have
\begin{equation}
\label{eq:consolidated}
H_\mathrm{TOT} = H_\mathrm{dev} + H_\mathrm{heat} + H_\mathrm{work} \, .
\end{equation}
We interpret the three terms on the right side of Eq.~\eqref{eq:consolidated} to be the instantaneous energies of the device, heat source and work source, respectively.
In our accounting, all interaction terms contribute to the energy of the device.

The microscopic evolution of our composite system is described by a Hamiltonian trajectory $\gamma(t)$, along which the value of $H_\mathrm{TOT}$ remains constant:
\begin{equation}
\label{eq:energyConservation}
\frac{\td }{\td t}\,  H_\mrm{TOT}\bigl(\gamma(t)\bigr)= \frac{\td  H_\mrm{dev}}{\td t} + \frac{\td H_\mrm{work}}{\td t} +\frac{\td H_\mrm{heat}}{\td t}=0 \, .
\end{equation}
The three subsystems exchange energy among themselves, with the total energy remaining fixed.
The heat and work sources are not directly coupled to one another, but each is coupled to the device.
Therefore the rate at which the work source loses energy is interpreted as the rate at which work is performed on the device:
\begin{equation}
\label{eq:defWork}
\frac{\td W}{\td t} \equiv - \frac{\td H_\mrm{work}}{\td t} \, .
\end{equation}
Similarly, energy lost by the heat source is equated with heat absorbed by the device:
\begin{equation}
\label{eq:defHeat}
\frac{\td Q}{\td t} \equiv - \frac{\td H_\mrm{heat}}{\td t} \, .
\end{equation}
Combining Eqs.~\eqref{eq:energyConservation}-\eqref{eq:defHeat}, we arrive at
\begin{equation}
\label{eq:firstLaw}
\dot H_\mrm{dev} = \dot W + \dot Q \, ,
\end{equation}
where the dots indicate derivatives with respect to time.
Equation~\eqref{eq:firstLaw} constitutes the first law of thermodynamics, in our framework.

By direct evaluation -- using Hamilton's equation for the work source, $\dot X = \partial H_\mathrm{TOT}/\partial P$, $\dot P = -\partial H_\mathrm{TOT}/\partial X$, together with Eq.~\eqref{e02} --
we obtain
\begin{equation}
\label{e04}
\frac{\td }{\td t}\,H_\mathrm{work}(X,P) = \dot X \frac{\partial H_\mrm{work}}{\partial X} + \dot P \frac{\partial H_\mrm{work}}{\partial P} =-\dot X \frac{\pd H_0}{\pd X}  \, .
\end{equation}
Thus, with Eq.~\eqref{eq:defWork}, the work performed on the device from time $t_1$ to $t_2$ is given by
\begin{equation}
\label{eq:defTotalWork}
W = \int_{t_1}^{t_2} {\td t} \,\, \dot X  \, \frac{\pd H_0}{\pd X} \, ,
\end{equation}
which is the familiar integral of {\it displacement $\times$ force} used in thermodynamics \cite{landau_80}.

\subsection{Work reservoir}
\label{subsec:work}

As mentioned earlier, there are two assumptions one might make about the properties of the work source: rapid self-equilibration and large inertia.
We have built the first assumption into our framework, by modeling the work source with a single degree of freedom, $X$.
We will now also make the second assumption, as this will allow us to address cyclic processes.

\begin{figure}
\centering
 \includegraphics[width=0.48\textwidth]{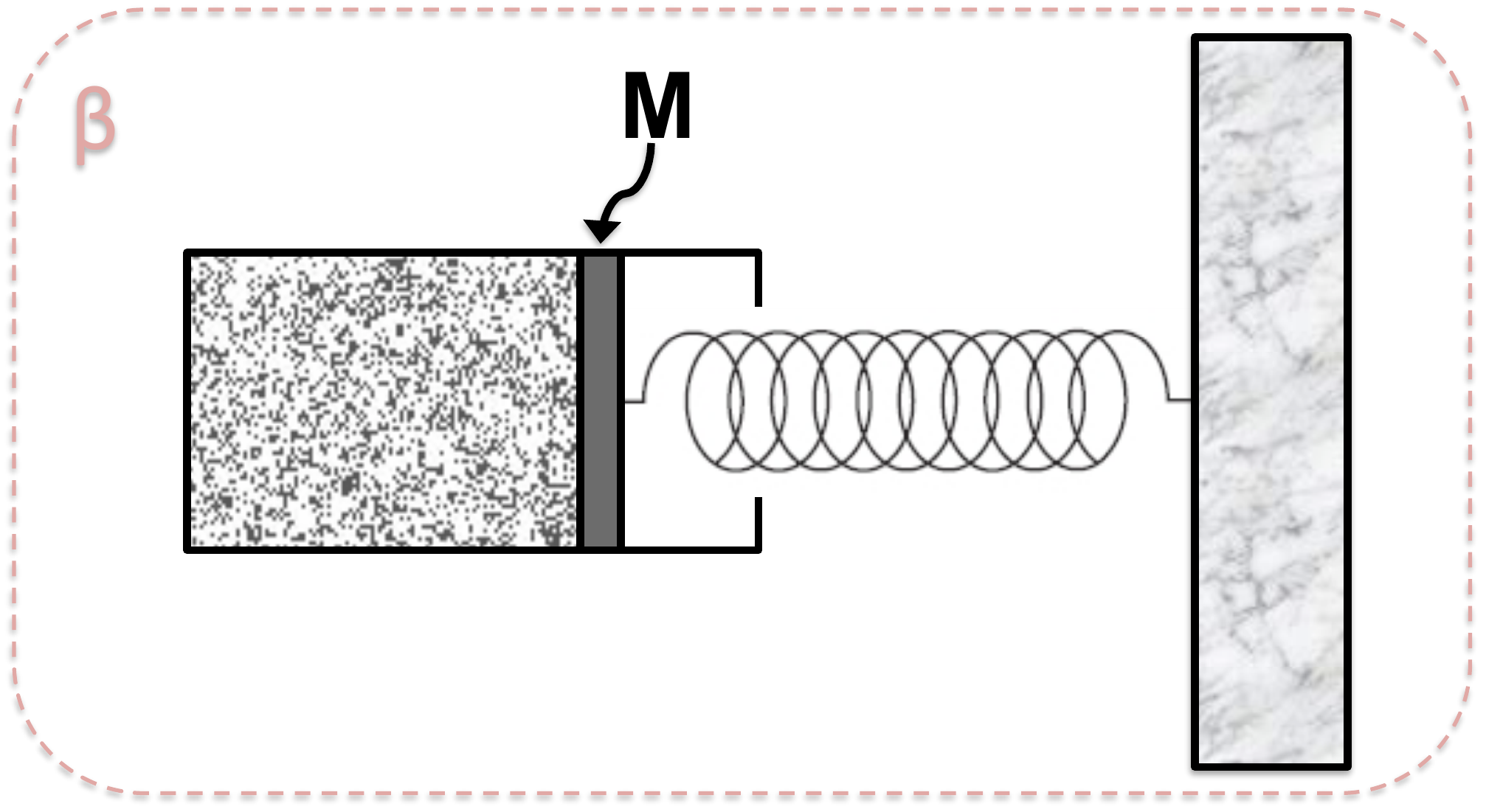}
\caption{\label{work_res} (color online) Example of a work reservoir: Spring is attached to a piston with large mass $M$ and confining a rarefied gas in a cylinder. The gas is in thermal contact with a heat bath of temperature $\beta^{-1}$.}
\end{figure}

To formalize the assumption of large inertia, let us consider a specific example, in which a massive piston (the work source) confines a rarefied gas (the device) within a cylinder.
The bare Hamiltonian for the work source is
\begin{equation}
H_\mathrm{work}(X,P)=\frac{P^2}{2M}+ \frac{M\omega^2 X^2}{2} \, ,
\end{equation}
where the potential energy term models an ideal spring attached to the piston, as illustrated in Fig.~\ref{work_res}.
The piston begins in a microstate $(X_0,P_0)$, then evolves together with the gas and the surrounding thermal reservoir over a time interval $0<t<t_f$, where $t_f$ specifies the duration of the process in which we are interested.

In the limit $M\rightarrow \infty$, with initial conditions $(\mb{x}_0,\mb{p}_0;\mb{\xi}_0,\mb{\varphi}_0;X_0,P_0/M)$ held fixed, the motion of the massive piston becomes unaffected by the remaining degrees of freedom, and is given by its free dynamics,
\begin{subequations}
\label{e08}
\begin{align}
\label{eq:Xt}
\lim_{M\rightarrow\infty} X(t)&= X_0 \co{\omega t} -\frac{V_0}{\omega} \si{\omega t}\\
\label{eq:Pt}
\lim_{M\rightarrow\infty} V(t)&= V_0 \co{\omega t}+X_0\,\omega\si{\omega t}\,,
\end{align}
\end{subequations}
where $V=P/M$ is the piston speed.
(See Appendix \ref{app:work} for details.)
Thus for sufficiently large $M$ we can treat the motion of the piston as fully prescribed, given the initial conditions.
This allows us to simplify the description of the total system.
The piston now evolves independently, and the device and heat source evolve under a Hamiltonian with an externally imposed time-dependence, determined by $X(t)$.
Introducing the notation
\begin{equation}
\begin{split}
& H_0(\mb{x},\mb{p};\, t) = H_0(\mb{x},\mb{p};\, X(t)) \\
& H_\mrm{dev}(\mb{x},\mb{p};\,\mb{\xi},\mb{\varphi};\, t) = H_\mrm{dev}(\mb{x},\mb{p};\,\mb{\xi},\mb{\varphi}; X(t)) \, ,
\end{split}
\end{equation}
with $X(t)$ given by Eq.~\eqref{eq:Xt}, we now define
\begin{equation}
\label{e09}
\begin{split}
H_\mathrm{tot}(\zeta;\, t) & \equiv H_\mrm{dev}(\mb{x},\mb{p};\,\mb{\xi},\mb{\varphi};\, t)+H_\mathrm{heat}(\mb{\xi},\mb{\varphi})  \\
& = H_0(\mb{x},\mb{p};\, t) + h(\mb{x},\mb{p};\,\mb{\xi},\mb{\varphi}) + H_\mathrm{heat}(\mb{\xi},\mb{\varphi}) \, ,
\end{split}
\end{equation}
where $\zeta = (\mb{x},\mb{p};\,\mb{\xi},\mb{\varphi})$ specifies a point in the reduced phase space of the device and heat source.
The Hamiltonian $H_\mrm{tot}$ gives the combined energy of these two subsystems, and generates their motion via Hamilton's equations.
Because it is explicitly time-dependent, its value is not preserved.
Rather, the net change in $H_\mrm{tot}$ along a trajectory $\zeta(t)$ corresponds to the work performed on the device.
This follows from energy conservation in the full phase space ($\dot H_\mrm{tot} = \dot H_\mrm{dev} + \dot H_\mrm{heat} = -\dot H_\mrm{work} = \dot W$, see Eqs.~\eqref{eq:energyConservation}, \eqref{eq:defWork}), as well as directly from Eq.~\eqref{eq:defTotalWork}:
\begin{equation}
\begin{split}
W &= \int_{t_1}^{t_2} {\td t} \, \dot X  \, \frac{\pd H_0}{\pd X}  \xrightarrow{M\rightarrow\infty}
\int_{t_1}^{t_2} {\td t} \, \frac{\pd H_0}{\pd t}
= \int_{t_1}^{t_2} {\td t} \, \frac{\pd H_\mrm{tot}}{\pd t} \\
&= H_\mrm{tot}\bigl(\zeta(t_2);t_2) - H_\mrm{tot}\bigl(\zeta(t_1);t_1)  \, .
\end{split}
\end{equation}
Here we have made use of the Hamiltonian identity ${\td H}_\mrm{tot}/{\td t} = {\pd H}_\mrm{tot}/{\pd t}$ \cite{goldstein_80}.

By construction, $H_\mrm{tot}(\zeta;t)$ is a periodic function of time, with period $\tau = 2\pi/\omega$.
The limit of large inertia, Eq.~\eqref{e08}, thus takes us from an inclusive description involving three subsystems (the device, the heat source and the work source) to a reduced description in which the device, coupled to the heat source, is subjected to time-periodic external driving.
We will continue our analysis within the reduced framework, making use of time-periodic Hamiltonians of the form given by Eq.~\eqref{e09}.
However, we emphasize that the explicit time-dependence of $H_\mrm{tot}$ is entirely induced by the dynamics of the massive work source.

We have used the piston-and-spring as an illustrative example, but the work source can equally well be modeled using a generic one-dimensional potential, provided the limit $M\rightarrow\infty$ is taken (as above) with $P_0/M$ held fixed.
Thus the time-dependence of the coordinate $X(t)$, while periodic, need not be sinusoidal.
In the remainder of the paper we will use $\tau$ to denote the period of the motion of the coordinate $X$ (in the large-inertia limit), whether it is harmonic or not.

\subsection{Heat reservoir}
\label{subsec:heat}

The limit of large work source inertia gives us time-periodic driving, as we have just argued, but does not yet guarantee that the device itself relaxes to a time-periodic steady state. For that we will require two assumptions about the heat source, namely that it is self-equilibrating and has a large heat capacity. In classical, macroscopic thermodynamics, these are among the defining properties of an idealized heat reservoir \cite{callen_85}. We now discuss these assumptions in the context of our explicitly microscopic setup, and we formulate a plausibility argument for the emergence of a periodic steady state, Eq.~\eqref{eq:pss}.

A large heat capacity implies that the number of degrees of freedom of our heat source, $N_\mrm{heat}$, far exceeds that of the device. This can be formalized by considering the thermodynamic limit, $N_\mrm{heat}\rightarrow\infty$, while holding fixed the intensive properties of the heat source -- its temperature, density, and chemical composition. In this limit, the characteristic energy exchanged between the device and the heat source, during the process in question, becomes a negligible fraction of the total energy of the heat source. Therefore its intensive properties, and particularly its temperature, remain unchanged.

We take the assumption of self-equilibration to mean the following: from a generic initial microstate and in the absence of external influences, the heat source evolves to a microstate that -- for purpose of subsequent calculations -- can be treated as a random sample from an equilibrium probability distribution~\footnote[1]{
For purpose of the present discussion, in the thermodynamic limit, it is not particularly relevant whether we view the equilibrium distribution to be microcanonical or canonical.
}.
A first-principles justification of this assumption involves issues that are well beyond the scope of this paper~\cite{khinchin49,dorfman99}.
Empirically, however, macroscopic systems do relax to equilibrium when left undisturbed (leaving aside special cases such as glassy systems), and these equilibrium states are accurately modeled by the standard probability distributions of classical statistical mechanics.
We will therefore assume that the heat source satisfies the property of self-equilibration, and we will investigate the consequences of this assumption.

Let us first consider the extreme limit, in which relaxation to equilibrium occurs on a time scale that is much faster than any other relevant time scale in our problem. In this case, even when the heat source interacts and exchanges energy with the device, its microstate at any instant can be treated as a random sample from an equilibrium ensemble. Effectively, then, the heat source evolves through a sequence of equilibrium states, as it absorbs or releases energy. Moreover, in the limit of infinite heat capacity, its temperature remains constant, $\beta_\tau\xrightarrow{N_\mrm{heat}\to\infty}\beta_0$, as discussed in a previous paragraph.

Now let $\mb{z}_n = \left(\mb{x}(n\tau),\mb{p}(n\tau)\right)$ denote the microstate of the device at the start of the $n$th period, and similarly define $\mb{Z}_n = \left(\mb{\xi}(n\tau),\mb{\varphi}(n\tau)\right)$ for the heat source.
The evolution of the combined system from one period to the next is given by the iteration of a deterministic mapping:
\begin{equation}
\label{eq:deterministicMapping}
\cdots \rightarrow
(\mb{z}_{n-1};\mb{Z}_{n-1}) \rightarrow
(\mb{z}_n;\mb{Z}_n) \rightarrow
(\mb{z}_{n+1};\mb{Z}_{n+1}) \rightarrow \cdots
\end{equation}
Each microstate $\zeta_n$ in this sequence is reached from the previous one, by evolving under Hamilton's equations for one period of the time-dependent Hamiltonian $H_{\rm tot}(\zeta;t)$.
In the limit of extremely rapid self-equilibration of the heat source, the $\mb{Z}_n$'s effectively become uncorrelated random samples from a fixed equilibrium distribution.
Abstractly, we can view ${\mb Z}_n$ as a set of freshly generated random numbers that collectively determine the value of ${\mb z}_{n+1}$, given ${\mb z}_n$;
in the next iteration, a new set of random numbers, ${\mb Z}_{n+1}$, determine the transition from ${\mb z}_{n+1}$ to ${\mb z}_{n+2}$, and so forth. 
Adopting this perspective, the stroboscopic evolution of the device from one period to the next,
\begin{equation}
\label{eq:MarkovChain}
\cdots \rightarrow
\mb{z}_{n-1} \rightarrow
\mb{z}_n \rightarrow
\mb{z}_{n+1} \rightarrow \cdots
\end{equation}
is given by the iteration of a stationary, stochastic, Markovian mapping.
This Markov chain relaxes to a unique stationary state described by a fixed, generally nonequilibrium distribution $\bar\rho_{\rm dev}(\mb{x},\mb{p})$.
(This is a consequence of the Perron-Frobenius theorem \cite{kampen_92}, under standard assumptions.)
Therefore the time-dependent probability distribution for the device relaxes to a periodic steady state,
\begin{equation}
\label{eq:pss}
\rho_{\rm dev}(\mb{x},\mb{p},t+\tau)=\rho_{\rm dev}(\mb{x},\mb{p},t) \, ,
\end{equation}
where $\rho_{\rm dev}(\mb{x},\mb{p},n\tau) = \bar\rho_{\rm dev}(\mb{x},\mb{p})$.

To reach Eq.~\eqref{eq:pss} we have assumed that the self-equilibration of the heat source occurs, in effect, infinitely rapidly.
Now we loosen this assumption by allowing the relaxation time scale of the heat source to be comparable to other time scales in the problem.
In this case ${\mb Z}_{n+1}$ in Eq.~\eqref{eq:deterministicMapping} may be statistically correlated with ${\mb Z}_n$ and with ${\mb z}_n$.
Nevertheless it is reasonable to assume that there exists some integer $K>0$, such that ${\mb Z}_{n+K}$ is statistically uncorrelated with ${\mb Z}_n$ and ${\mb z}_n$.
In other words, a time interval of duration $K\tau$ is sufficient for the heat source to ``forget'' its microstate.
Then the stroboscopic evolution of the device in time increments $K\tau$,
\begin{equation}
\label{eq:KMarkovChain}
\cdots \rightarrow
\mb{z}_{n-K} \rightarrow
\mb{z}_n \rightarrow
\mb{z}_{n+K} \rightarrow \cdots
\end{equation}
is a Markov chain.
The ${\bm Z}_n$'s are no longer necessarily sampled from equilibrium.
However, if the heat source itself reaches a stationary state, in which the energy exchanged with the device is transported at a fixed rate to more distant regions of the heat source, then Eq.~\eqref{eq:KMarkovChain} becomes a stationary Markov chain, and the final arguments of the previous paragraph continue to apply: the device eventually relaxes to a periodic steady state.

As mentioned, the reasoning of the preceding paragraphs is intended as a plausibility argument for the emergence of cyclic motion of the device, under the conditions and limits we have discussed: the large inertia of the work source induces a time-periodic Hamiltonian for the device, and the large heat capacity and self-equilibration of the heat source cause the relaxation of the device into a time-periodic steady state.
For the remainder of this paper we will assume that these arguments apply -- hence the device reaches a periodic steady state -- and we will explore their consequences.
As suggested at the beginning of Section~\ref{sec:framework} we will henceforth use the terms {\it work reservoir} and {\it heat reservoir}.

We note that once the device has reached a periodic steady state, both its internal energy and its Shannon entropy become time-periodic as well:
\begin{equation}
\label{eq:pss_entropy_energy}
\langle H_\mrm{dev}(t+\tau)\rangle = \langle H_\mrm{dev}(t)\rangle
\quad , \quad
\mc{H}_\mrm{dev}(t+\tau)=\mc{H}_\mrm{dev}(t) \, .
\end{equation}
We will make use of this observation in our later analysis.

\subsection{Information reservoir}
\label{subsec:info}

In the preceding subsections, work and heat reservoirs have been discussed within a classical, Hamiltonian framework. We now complete this framework by introducing the possibility of information processing.  In effect, we aim to describe thermodynamic processes in the presence of a physical device capable of acting like Maxwell's demon, performing microscopic measurements and feedback on the other subsystems in our picture.  The key feature that we wish to capture is the demon's memory, where it stores information that it has gathered. To this end we introduce an idealized {\it information reservoir}, representing the demon's memory.  All other components of the mechanical demon are implicitly treated as belonging to the device of interest.

To describe the complete system consisting of device, heat reservoir, and information reservoir the total Hamiltonian \eqref{e09} is extended to read 
\begin{equation}
\label{e14}
\begin{split}
H_\mathrm{tot}(\zeta;\,\mb{\Xi},\mb{\Phi};\, t)&= H_\mrm{dev}(\mb{x},\mb{p};\,\mb{\xi},\mb{\varphi};\,t)+H_\mathrm{heat}(\mb{\xi},\mb{\varphi})\\
&+ H_\mrm{info}(\mb{\Xi},\mb{\Phi})+\mf{h}(\zeta;\,\mb{\Xi},\mb{\Phi})\,,
\end{split}
\end{equation}
where $(\mb{\Xi},\mb{\Phi})$ is the microstate of the information reservoir and $H_\mrm{info}$ is its bare Hamiltonian. The term $\mf{h}(\zeta;\,\mb{\Xi},\mb{\Phi})$ describes the interaction between the information reservoir and the device and thermal reservoir. The assumption that the information and thermal reservoirs are coupled is important for the following discussion.

Let us first describe the information reservoir in the presence of a single thermal reservoir, at inverse temperature $\beta$, before discussing its interaction with the device.

For specificity, we will take the information reservoir to be a memory register consisting of $N$ bits~\footnote[2]{
In general, of course, information can be stored by other means, for instance using {\it trits} -- trinary digits -- rather than bits. Our choice of using bits is motivated by convenience and familiarity, and does not restrict the validity of our conclusions.}.
A single bit is physically implemented using a large collection of atoms or molecules, whose total magnetization (or some other collective observable) acts as a binary order parameter.
We will distinguish between the {\it microstate} of the information reservoir, $\psi \equiv (\mb{\Xi},\mb{\Phi})$, and its {\it informational state}, $\sigma$.
The microstate $\psi$ is a point in the phase space of the entire collection of atoms and molecules comprising the memory register, whereas the informational state $\sigma$ is a given sequence of bit values, e.g.\ ${\tt 0110 \cdots 10}$.
We assume that each microstate $\psi$ corresponds to a particular informational state $\sigma$, and we will use the function $\hat\sigma(\psi)$ to specify the informational state associated with the microstate $\psi$.
The variables $\psi$ and $\sigma$ thus represent fine-grained and coarse-grained descriptions of the state of the information reservoir.

The function $\hat\sigma(\psi)$ partitions the phase space of the information reservoir into $2^N$ distinct regions, each corresponding to one informational state.
To guarantee a stable and reliable memory register, we assume these regions are separated by large free-energetic barriers, so that over the time scales that concern us, the probability of a spontaneous, thermally driven transition from one informational state to another is negligible.
It then becomes useful to consider a {\it constrained} equilibrium state, described by a conditional probability distribution
\begin{equation}
\label{eq:constrainedEqstate}
p^{\rm eq}(\psi\vert\sigma) = \delta_{\sigma,\hat\sigma(\psi)} \, \e{-\beta [H_{\rm info}(\psi) - F_{\rm info}^\sigma]} \, .
\end{equation}
Here, the Kronecker $\delta$-function acts as an indicator variable, hence $p^{\rm eq}(\psi\vert\sigma)$ is simply a canonical probability distribution, restricted to the region of phase space corresponding to the information state $\sigma$.
The free energy $F_{\rm info}^\sigma$ is determined by normalization, $\int {\rm d}\psi \, p^{\rm eq}(\psi\vert\sigma) = 1$, and in the usual manner we can define an equilibrium internal energy and entropy:
\begin{equation}
\label{eq:constrainedEnergyEntropy}
\begin{split}
\langle H_{\rm info}\rangle^{{\rm eq},\sigma} &= \int {\rm d}\psi \, p^{\rm eq}(\psi\vert\sigma) \, H_{\rm info}(\psi) \\
\mc{H}_{\rm info}^{{\rm eq},\sigma} &= - \int {\rm d}\psi \, p^{\rm eq}(\psi\vert\sigma) \ln p^{\rm eq}(\psi\vert\sigma)
\end{split}
\end{equation}
Equation~\eqref{eq:constrainedEqstate} represents the statistical state of the information reservoir, when it has been left undisturbed in the informational state $\sigma$, in the presence of a thermal reservoir.

Following Bennett~\cite{ben03}, we will refer to the $N$ bits as {\it information bearing degrees of freedom}, or {\it IBD}, and the remaining microscopic variables as {\it non-information bearing degrees of freedom}, or {\it NBD}.
Using this terminology, Eq.~\eqref{eq:constrainedEqstate} represents an equilibrium state of the NBD ($\psi\vert\sigma$), for a given state of the IBD ($\sigma$).

Let us now consider the behavior of the information reservoir in the presence of the device of interest.
We explicitly assume that interactions with the device can give rise to transitions among the informational states.
In this manner, information about the evolution of the device of interest becomes encoded in the IBD.
Let us further assume that:
(1) the $2^N$ informational states have the same equilibrium energies and entropies,
and
(2) after a transition from one informational state to another, thermal equilibration of the NBD occurs rapidly.
Under these assumptions, the energy of the information reservoir effectively remains constant, aside from equilibrium thermal fluctuations. In the presence of the device of interest and thermal reservoir, the evolution of the information reservoir is a sequence of transitions from one equilibrated informational state to another.

The total information encoded in the reservoir is quantified by its Shannon entropy, which can formally be decomposed into contributions from the information bearing and non-information bearing degrees of freedom:
\begin{equation}
\label{eq:decomposeHinfo}
\mc{H}_\mrm{info}(t) = -\tr{\rho_\mrm{info}(t)\ln\rho_\mrm{info}(t)}
= \mc{H}_\mrm{info}^\mrm{IBD}(t) + \mc{H}_\mrm{info}^\mrm{NBD}(t) \, ,
\end{equation}
as we show in Appendix \ref{app:info}.
Moreover, under the assumptions of the previous paragraph, $\mc{H}_\mrm{info}^\mrm{NBD}(t)$ does not vary with time, and is given simply by the equilibrium entropy of the microscopic, non-information bearing degrees of freedom (again, see Appendix \ref{app:info} for details).
As a result, any \textit{change} in the Shannon entropy of the information reservoir, resulting from its interactions with the device over an interval of time, is entirely captured by the net change in the probability distribution of the mesoscopic, information-bearing degrees of freedom:
\begin{equation}
\label{eq:deltaHinfo}
\Delta \mc{H}_\mrm{info} = \Delta \mc{H}_\mrm{info}^\mrm{IBD} \, .
\end{equation}
In Sections \ref{sec:ft} - \ref{sec:maxwork} below, we will use the notation $\Delta \mc{H}_\mrm{info}$ rather than $\Delta \mc{H}_\mrm{info}^\mrm{IBD}$, to avoid clutter, but it will be understood that the net change in the Shannon entropy of the information reservoir refers to the change in its information-bearing degrees of freedom.

\section{Non-negativity of information exchange}
\label{sec:ft}

The rest of this paper is devoted to investigating specific thermodynamic processes within the framework introduced above.  To this end, we begin by obtaining an inequality for the sum of changes of the Shannon entropy for the individual subsystems, Eq.~\eqref{e18}, from which we derive an inequality related to the behavior of our system in the periodic steady state, Eq.~\eqref{e18e}.  The latter result, and its generalization, Eq.~\eqref{e18g}, will then by exploited in Section~\ref{sec:2ndlaw}.  As in the recent work of Hasegawa {\it et al}~\cite{has10,tak10} and Esposito {\it et al}~\cite{esp10a,esp11}, our approach in this section will draw on properties of the canonical distribution, the Shannon entropy, and the Kullback-Leibler divergence~\cite{kullback_78}, as well as assumptions about the initial state of the system.

We adopt an explicitly statistical perspective, in which we consider an {\it ensemble} representing different possible microscopic realizations of the process. The probability distribution in the full phase space at an initial time $t=0$ reflects the preparation of the system prior to this time, and we now spell out the assumptions that we make regarding this preparation. As in Refs.~\cite{jar99,pie00,has10,esp10a,tak10,esp11}, we assume that the total system begins in a product state,
\begin{equation}
\begin{split}
\label{eq:initial_state}
\rho_\mrm{tot}(\zeta;\,\mb{\Xi},\mb{\Phi};\,0)&=\rho_\mrm{dev}(\mb{x},\mb{p};\,0)\times\rho_\mrm{heat}(\mb{\xi},\mb{\varphi};\,0)\\
&\times\rho_\mrm{info}(\mb{\Xi},\mb{\Phi};\,0)\,.
\end{split}
\end{equation}
This assumption does not substantially restrict the generality of the following discussion, as we expect the device to relax into a time-periodic steady state that is independent of its initial preparation (see Section~\ref{sec:framework}).  For the time being, we restrict ourselves to considering only a single thermal reservoir, but as discussed below the results generalize easily to multiple reservoirs; see e.g.\ Eq.~\eqref{e18g}.

We take the initial state of the heat reservoir to be given by the canonical distribution,
\begin{equation}
\label{eq:heat_reservoir_equilibrium}
\rho_\mrm{heat}(\mb{\xi},\mb{\varphi};\,0) = \frac{1}{Z_\mrm{heat}} \exp \left[ -\beta H_\mrm{heat}(\mb{\xi},\mb{\varphi}) \right] \equiv \rho_\mrm{heat}^\mrm{eq}(\mb{\xi},\mb{\varphi}) \, ,
\end{equation}
with free energy $F_\mrm{heat} = -\beta^{-1} \ln Z_\mrm{heat}$.
For the distribution $\rho_\mrm{info}(\mb{\Xi},\mb{\Phi};\,0)$, we assume that the microscopic, non-information bearing degrees of freedom are in equilibrium with the thermal reservoir (see Section \ref{subsec:info}), whereas the distribution of the mesoscopic, information bearing degrees of freedom reflects the manner in which the information reservoir was prepared.
For instance, the memory register might be initialized in a \textit{blank} state, {\tt 000$\cdots$0}, in which case $\mc{H}_\mrm{info}^\mrm{IBD} = 0$.
At the other extreme, it may be prepared so that every possible $N$-bit sequence (informational state) is equally likely, hence $\mc{H}_\mrm{info}^\mrm{IBD} = N \ln 2$.
We will not place any restrictions on the initial statistical state of the device, $\rho_\mrm{dev}(\mb{x},\mb{p};\,0)$.

After the initial preparation the full system evolves under the time-periodic Hamiltonian given by Eq.~\eqref{e14}. In general, the total density at time $t>0$, $\rho_\mrm{tot}(\zeta;\,\mb{\Xi},\mb{\Phi};\,t)$, will not be a product state and the reduced densities for device, heat reservoir, and memory are obtained by integrating out the other subsystems. Thus for the device we have
\begin{equation}
\label{e16}
\rho_\mrm{dev}(\mb{x},\mb{p};\,t)=\int \td\mb{\xi}\,\td\mb{\varphi}\int \td\mb{\Xi}\,\td\mb{\Phi}\,\rho_\mrm{tot}(\zeta;\,\mb{\Xi},\mb{\Phi};\,t)\,,
\end{equation} 
and similarly for the heat and information reservoir.
We can use these reduced densities to define the Shannon entropy \cite{cover_91} of each subsystem, e.g.\
\begin{equation}
\begin{split}
\mc{H}_\mrm{dev}(t) &=-\tr{\rho_\mrm{dev}(t)\ln\rho_\mrm{dev}(t)} \\
&\equiv - \int \td\mb{x}\,\td\mb{p} \,\, \rho_\mrm{dev}(\mb{x},\mb{p};\,t) \ln \rho_\mrm{dev}(\mb{x},\mb{p};\,t)
\, ,
\end{split}
\end{equation}
and analogously for the heat and information reservoir.

By Liouville's theorem, the Shannon entropy of the total system remains constant under Hamiltonian dynamics: $\mc{H}_\mrm{tot}(t)=\mc{H}_\mrm{tot}(0)$.  Moreover, since the system is prepared in a product state \eqref{eq:initial_state}, we have
\begin{equation}
\label{eq:initial_Shannon}
\mc{H}_\mrm{tot}(0)=\mc{H}_\mrm{dev}(0)+\mc{H}_\mrm{heat}(0)+\mc{H}_\mrm{info}(0)\,.
\end{equation}
At later times $t$ we have
\begin{equation}
\label{eq:final_Shannon}
\mc{H}_\mrm{tot}(t)\leq\mc{H}_\mrm{dev}(t)+\mc{H}_\mrm{heat}(t)+\mc{H}_\mrm{info}(t)\,,
\end{equation}
due to the subadditivity of the Shannon entropy \cite{cover_91}. Subtracting Eq.~\eqref{eq:initial_Shannon} from Eq.~\eqref{eq:final_Shannon} we obtain
\begin{equation}
\label{e18}
\Delta\mc{H}_\mrm{dev}+\Delta\mc{H}_\mrm{heat}+\Delta\mc{H}_\mrm{info}\geq \Delta\mc{H}_\mrm{tot}=0\, ,
\end{equation}
where $\Delta\mc{H}_\mrm{dev}$ denotes the net change in the Shannon entropy of the device of interest, and similarly for the other subsystems.

For the heat reservoir we can write
\begin{equation}
\label{e18a}
\begin{split}
\mc{H}_\mrm{heat}(t)&=-\tr{\rho_\mrm{heat}(t)\ln{\rho_\mrm{heat}(t)}}\\
&=\beta E_\mrm{heat}(t) -\beta F_\mrm{heat}-D\left(\rho_\mrm{heat}(t)||\rho_\mrm{heat}^\mrm{eq}\right)\,,
\end{split}
\end{equation}
where
\begin{equation}
E_\mrm{heat}(t) \equiv \tr{ H_\mrm{heat}\, \rho_\mrm{heat}(t)}
\end{equation}
is the average energy of the reservoir, and $D(\cdot || \cdot)$ denotes the Kullback-Leibler divergence~\cite{kullback_78}
\begin{equation}
\label{eq:KLdef}
D(\rho||\sigma)=\tr{\rho\ln\rho}-\tr{\rho\ln\sigma}\geq 0 \, .
\end{equation}
Combining Eqs.~\eqref{e18a} with our assumption that the heat reservoir is prepared in equilibrium, Eq.~\eqref{eq:heat_reservoir_equilibrium}, we get
\begin{equation}
\label{e18b}
\begin{split}
\Delta \mc{H}_\mrm{heat}&=\beta E_\mrm{heat}(t)-\beta E_\mrm{heat}(0)-D\left(\rho_\mrm{heat}(t)||\rho_\mrm{heat}^\mrm{eq}\right)\\
&\leq\beta \left[E_\mrm{heat}(t)- E_\mrm{heat}(0)\right] = \beta \Delta E_\mrm{heat} \, ,
\end{split}
\end{equation}
using Eq.~\eqref{eq:KLdef}. This result in turn combines with Eq.~\eqref{e18} to give
\begin{equation}
\label{e18c}
\Delta\mc{H}_\mrm{dev}+\beta \Delta E_\mrm{heat}+\Delta\mc{H}_\mrm{info}\geq 0\,.
\end{equation}

Note that we have taken two distinct steps to arrive at Eq.~\eqref{e18c}.
First, we have obtained Eq.~\eqref{e18} from our assumption that the subsystems are statistically uncorrelated at the initial time, Eq.~\eqref{eq:initial_state}.
In fact, the left side of Eq.~\eqref{e18} quantifies the degree to which correlations develop between the subsystems, due to their mutual interactions;
Esposito {\it et al}~\cite{esp10a} have explicitly interpreted this build-up of correlations as representing entropy production.
Next, to get to Eq.~\eqref{e18c} we have used the assumption that the reservoir is initialized in the canonical distribution, Eq.~\eqref{eq:heat_reservoir_equilibrium}, together with the non-negativity of the Kullback-Leibler divergence.
Similar manipulations appear in Refs.~\cite{has10,esp10a,tak10,esp11}.
We will now use Eq.~\eqref{e18c} to arrive at inequalities that characterize the behavior of our system in the periodic steady state.

A natural time scale for our process is given by the driving period $\tau$ (Section~\ref{subsec:work}). Let us set $t = n\tau > n_0\tau$, where $n_0$ is the number of periods needed for the device to relax into its periodic steady state.
Then the process in question can be divided into a transient interval ($0\rightarrow n_0\tau$) followed by an interval of time-periodic behavior ($n_0\tau \rightarrow n\tau$).
Expressing each term in Eq.~\eqref{e18c} as a sum of contributions from these two intervals, we get
\begin{equation}
\begin{split}
&\Delta\mc{H}_\mrm{dev}^{0\rightarrow n_0}+\beta \Delta E_\mrm{heat}^{0\rightarrow n_0}+\Delta\mc{H}_\mrm{info}^{0\rightarrow n_0}\\
&+\Delta\mc{H}_\mrm{dev}^{n_0\rightarrow n}+\beta \Delta E_\mrm{heat}^{n_0\rightarrow n}+\Delta\mc{H}_\mrm{info}^{n_0\rightarrow n} \geq 0 \, ,
\end{split}
\end{equation}
which can further be rewritten as
\begin{equation}
\label{e18d}
\begin{split}
&\Delta\mc{H}_\mrm{dev}^{0\rightarrow n_0}+\beta \Delta E_\mrm{heat}^{0\rightarrow n_0}+\Delta\mc{H}_\mrm{info}^{0\rightarrow n_0}\\
&+(n-n_0)\,\left(\beta \Delta E_\mrm{heat}^\mrm{cyc}+\Delta\mc{H}_\mrm{info}^\mrm{cyc}\right)\geq 0\, ,
\end{split}
\end{equation}
where
\begin{equation}
\Delta E_\mrm{heat}^\mrm{cyc} = \frac{\Delta E_\mrm{heat}^{n_0\rightarrow n}}{n-n_0} 
\end{equation}
is the average heat absorbed by the heat reservoir, per cycle, in the periodic steady state, and
\begin{equation}
\Delta\mc{H}_\mrm{info}^\mrm{cyc} = \frac{\Delta\mc{H}_\mrm{info}^{n_0\rightarrow n}}{n-n_0}
\end{equation}
is the average change in the Shannon entropy of the information reservoir, per cycle, in the periodic steady state.
Note that the similarly defined quantity $\Delta\mc{H}_\mrm{dev}^\mrm{cyc}$ vanishes, by Eq.~\eqref{eq:pss_entropy_energy}.
Dividing both sides of the inequality in Eq.~\eqref{e18d} by $(n-n_0)$, then taking the limit $n\rightarrow\infty$, we finally obtain
\begin{equation}
\label{e18e}
\beta \Delta E_\mrm{heat}^\mrm{cyc}+\Delta\mc{H}_\mrm{info}^\mrm{cyc}\geq 0\,.
\end{equation}
In Section~\ref{sec:2ndlaw} we will exploit this result (or its generalization, Eq.~\eqref{e18g}) to obtain generalized versions of the Kelvin-Planck, the Clausius, and the Carnot statements of the second law.

Equation~\eqref{e18e} has a simple interpretation: the first term on the left represents the net change in the thermodynamic entropy of the heat reservoir, and the second term is the net change in the Shannon entropy of the information reservoir, specifically its information-bearing degrees of freedom.
Either term can be positive, negative or zero, but their sum must be non-negative.

In the preceding analysis, for convenience, we have restricted ourselves to a single heat reservoir. The arguments are readily generalized to the case of multiple heat reservoirs, by replacing the change of Shannon entropy for one reservoir by a sum over all reservoirs, in Eq.~\eqref{e18}, and by assuming that each heat reservoir is independently prepared in a canonical distribution corresponding to a particular temperature.
In particular for one hot and one cold reservoir Eq.~\eqref{e18e} becomes
\begin{equation}
\label{e18g}
\beta_\mrm{hot}\, \Delta E_\mrm{hot}^\mrm{cyc}+\beta_\mrm{cold}\, \Delta E_\mrm{cold}^\mrm{cyc}+\Delta\mc{H}_\mrm{info}^\mrm{cyc}\geq 0\,.
\end{equation}
In the case of multiple reservoirs, we will assume that the information reservoir is coupled only to a single thermal reservoir, and its microscopic degrees of freedom remain in equilibrium at the corresponding temperature.
The results that we derive in the following section do not depend on which reservoir is selected for this role.

\section{The second law and information processing}
\label{sec:2ndlaw}

In the periodic steady state, $\la H_\mrm{dev}(t+\tau)\ra=\la H_\mrm{dev}(t)\ra$ (Eq.~\eqref{eq:pss_entropy_energy}). Therefore, by the first law of thermodynamics, integrating Eq.~\eqref{eq:firstLaw} over a single cycle, we have
\begin{equation}
\label{eq:1stlaw}
\la W^\mrm{cyc} \ra +\la Q^\mrm{cyc}\ra = 0 \,,
\end{equation}
where $W^\mrm{cyc}$ is the net work performed on the device, and $Q^\mrm{cyc}$ is the net heat absorbed by the device (from one or more heat reservoirs), over one cycle in the periodic steady state; and  angular brackets denote averages over many realizations. By Eq.~\eqref{eq:defHeat}, the heat absorbed by the system is defined as the net decrease in the bare energies of the heat reservoir(s), hence Eq.~\eqref{eq:1stlaw} becomes
\begin{equation}
\label{eq:1stlaw_again}
\la W^\mrm{cyc} \ra = \Delta E_\mrm{heat}^\mrm{cyc}  \,.
\end{equation}

\subparagraph{Kelvin-Planck statement}

The Kelvin-Planck statement \cite{thomson_82} expresses the observation that in cyclic, isothermal processes the average work is always non-negative, $\la W^\mrm{cyc}\ra \geq 0$. To generalize this statement we consider a device of interest, coupled to a single heat reservoir at inverse temperature $\beta$, a work reservoir, and an information reservoir. Equation~\eqref{eq:1stlaw_again} then combines with Eq.~\eqref{e18e} to give:
\begin{equation}
\label{eq:KP}
\beta \la W^\mrm{cyc}\ra\geq -\Delta\mc{H}_\mrm{info}^\mrm{cyc}\,,
\end{equation}
which constitutes a generalized version of the Kelvin-Planck statement.  For processes during which information is {\it written to} the information reservoir ($\Delta\mc{H}_\mathrm{info}^\mrm{cyc} > 0$) the net work over one cycle can be negative.  In other words, there can be a systematic transfer of energy from the heat reservoir to the work reservoir, provided the Shannon entropy of the information reservoir increases.  This is consistent with the current consensus regarding the Maxwell demon paradox~\cite{penrose_1970,ben82,leff_2003,man12}. For processes during which information is {\it erased} ($\Delta\mc{H}_\mathrm{info}^\mrm{cyc} < 0$) Eq.~\eqref{eq:KP} becomes equivalent to Landauer's principle~\cite{lan61,pie00,has10,tak10,esp11}, placing a lower limit on the amount of work that must be expended in order to accomplish this erasure.

\subparagraph{Clausius statement}

\begin{figure}
\centering
 \includegraphics[width=0.47\textwidth]{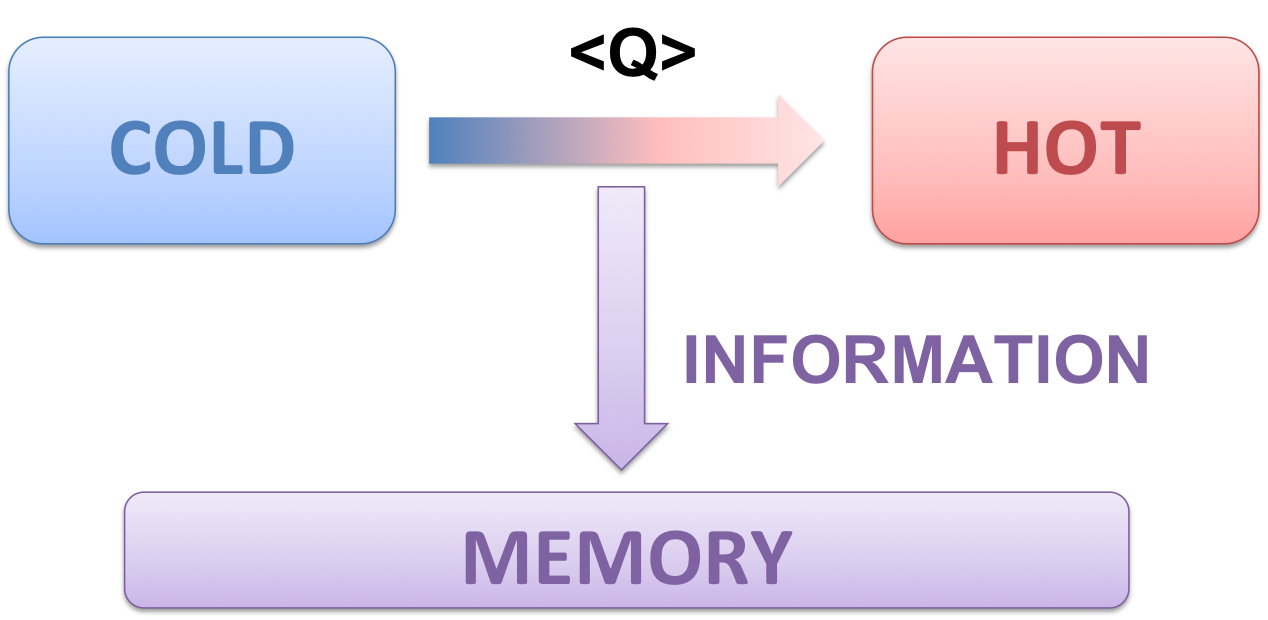}
\caption{\label{claus_info}Illustration of the generalized Clausius statment \eqref{eq:CL}. Heat can flow from the cold to the hot reservoir if information is written to memory.}
\end{figure}

To generalize the Clausius statement we consider a device interacting with two heat reservoirs, one hot and one cold, as well as a work reservoir and an information reservoir. As above, these interactions produce exchanges of both energy and information.
\begin{figure}
\centering
\includegraphics[width=0.47\textwidth]{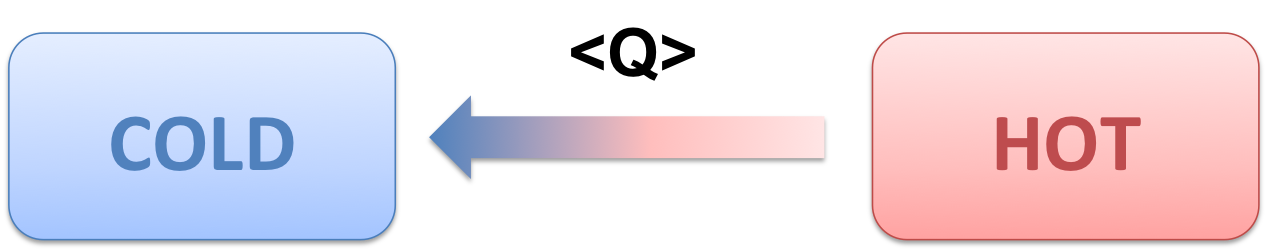}
\caption{\label{claus}Illustration of the Clausius statement. On average, heat always flows from the hot to the cold reservoir.}
\end{figure}
In general the net work performed on the device over one cycle can have either sign, and the device may be able to operate as either a heat engine or a refrigerator.  For the Clausius statement we restrict our attention to processes for which $\la W^\mrm{cyc}\ra=0$. In this case Eq.~\eqref{eq:1stlaw_again} becomes
\begin{equation}
0 = \Delta E_\mrm{hot}^\mrm{cyc} + \Delta E_\mrm{cold}^\mrm{cyc} \, .
\end{equation}
Consequently, Eq.~\eqref{e18g} can be written as
\begin{equation}
\label{eq:CL}
\begin{split}
\left(\beta_\mathrm{cold}-\beta_\mathrm{hot}\right) \la Q_\mrm{hot}^\mrm{cyc}\ra\geq-\Delta\mc{H}^\mrm{cyc}_\mathrm{info}\,,
\end{split}
\end{equation}
which generalizes the Clausius statement. Note that the left side of Eq.~\eqref{eq:CL} represents classical thermodynamic entropy, i.e. the heat exchanged over temperature, whereas the rights side quantifies the internal information gain in the memory.  Since $\beta_\mrm{cold} > \beta_\mrm{hot}$, Eq.~\eqref{eq:CL} allows for processes during which heat flows systematically from cold to hot ($\la Q_\mrm{hot}\ra < 0 < \la Q_\mrm{cold} \ra$) provided information is written to the memory, as illustrated schematically in Fig.~\ref{claus_info}. Conversely, if information is to be erased, then the right side of the inequality is positive, and we get a lower bound on the amount of heat that must flow from the hot to the cold reservoir.  For the erasure of one bit of information per cycle, $\Delta\mc{H}_\mathrm{info}^\mrm{cyc} = -\ln 2$, the average heat flow must satisfy
\begin{equation}
\la Q_\mrm{hot}^\mrm{cyc}\ra \ge (\beta_\mrm{cold}-\beta_\mrm{hot}) \ln 2 \, ,
\end{equation}
which represents a modified version of Landauer's principle~\cite{man13}.

\subparagraph{Carnot statement}

The Carnot statement asserts that the efficiency of a heat engine is always less than the Carnot efficiency, $\eta\leq\eta_\mathrm{C} \equiv 1-\beta_\mathrm{hot}/\beta_\mathrm{cold}$ \cite{carnot_24}. To generalize this result, we again consider a device interacting with two heat reservoirs, a work reservoir and an information reservoir, but now we consider processes for which the work performed on the device over one cycle is negative (in other words, the device delivers work) and the heat absorbed from the hot reservoir is positive, hence the devices operates as a heat engine, with efficiency $\eta=-\la W^\mrm{cyc}\ra/\la Q_\mathrm{hot}^\mrm{cyc}\ra > 0$. Equation \eqref{e18g} takes the form
\begin{equation}
\label{e28}
-\beta_\mrm{hot}\, \la Q_\mrm{hot}^\mrm{cyc}\ra-\beta_\mrm{cold}\, \la Q_\mrm{cold}^\mrm{cyc}\ra+\Delta\mc{H}_\mrm{info}^\mrm{cyc}\geq 0\, ,
\end{equation}
and Eq.~\eqref{eq:1stlaw} can be written as
\begin{equation}
\la Q_\mrm{cold}^\mrm{cyc}\ra=-\la W^\mrm{cyc} \ra -\la Q_\mrm{hot}^\mrm{cyc}\ra\,.
\end{equation}
Combining these equations we obtain
\begin{equation}
-\beta_\mrm{hot}\, \la Q_\mrm{hot}^\mrm{cyc}\ra+\beta_\mrm{cold}\, \la W^\mrm{cyc} \ra +\beta_\mrm{cold} \,\la Q_\mrm{hot}^\mrm{cyc}\ra +\Delta\mc{H}_\mrm{info}^\mrm{cyc}\geq 0\,.
\end{equation}
After rearrangement of terms, we find that the efficiency must satisfy
\begin{equation}
\label{eq:CA}
\eta\leq\left(1-\frac{\beta_\mathrm{hot}}{\beta_\mathrm{cold}}\right)+\frac{\Delta \mc{H}^\mrm{cyc}_\mathrm{info}}{\beta_\mathrm{cold}\la Q_\mathrm{hot}\ra}
 = \eta_\mathrm{C}+\frac{\Delta \mc{H}^\mrm{cyc}_\mathrm{info}}{\beta_\mathrm{cold}\la Q_\mathrm{hot}\ra}\,.
\end{equation}
Thus for cyclic processes in which information is systematically written to the memory, the efficiency can exceed the Carnot limit.
Note that Eq.~\eqref{eq:CA} does not depend on whether the information reservoir is coupled to the hot or the cold heat reservoir.

\section{Maximum work theorem}
\label{sec:maxwork}

In Section~\ref{sec:2ndlaw} we considered only cyclic processes.  Now let us briefly consider what happens when we relax this restriction. For \textit{non-cyclic} processes in the presence of a single heat reservoir, the second law is formulated in terms of the Helmholtz free energy, $F= E-\beta^{-1}S$, where $E=\la H\ra$ is the mean internal energy, $F$ the free energy, and $S$ the thermodynamic entropy of the system in question, in a state of thermal equilibrium. If the systems begins in one equilibrium state and ends in another, then the average work performed on the system during the process satisfies $\la W\ra\geq\Delta F$, where the equality holds for reversible processes.  Equivalently, the decrease in free energy gives the maximum usable, i.e. extractable work during such a process.

To generalize this result, we consider a device of interest, coupled to a single heat reservoir, a work reservoir and an information reservoir, without assuming cyclic motion.  As before, we assume an initial product state, Eq.~\eqref{eq:initial_state}, without imposing any restrictions on the initial state of the device; and we imagine observing the entire system over some interval of time.
Integrating Eq.~\eqref{eq:firstLaw} over this interval, we get 
\begin{equation}
\label{e32}
\Delta E_\mrm{dev} \equiv \la \Delta H_\mrm{dev} \ra =\la W\ra +\la Q\ra=\la W\ra-\Delta E_\mathrm{heat}\,,
\end{equation}
averaging over many realizations of the process. Combining this with Eq.~\eqref{e18c}, which was derived without assuming cyclic processes, we obtain
\begin{equation}
\label{e33}
\Delta\mc{H}_\mrm{dev}-\beta\Delta E_\mrm{dev}+\beta \la W\ra +\Delta\mc{H}_\mrm{info}\geq 0\,.
\end{equation}
In order to further simplify Eq.~\eqref{e33} we introduce the information free energy,
\begin{equation}
\label{eq:F}
\mc{F} = E_\mrm{dev} - \beta^{-1}\mc{H}_\mrm{dev}  = F + D(\rho||\rho^\mathrm{eq}) \, ,
\end{equation}
which generalizes the equilibrium free energy to an arbitrary non-equilibrium state characterized by a probability distribution $\rho$. This nonequilibrium free energy has previously appeared in both Hamiltonian treatments~\cite{has10,esp10a,tak10}, for instance to derive Landauer's principle \cite{esp11}, as well as stochastic treatments~\cite{qian_2001,crooks_2007,ge10,still_2012,sag12}. A generalized free energy of this form has also appeared in the thermodynamic description of open system dynamics~\cite{klimontovich_1999}. More recently, it was shown that $\mc{F}$ is a Lyapunov function for nonequilibrium stationary states \cite{def12}. 

In terms of this quantity, Eq.~\eqref{e33} becomes
\begin{equation}
\label{e34}
\beta \la W\ra\geq\beta\Delta \mc{F} -\Delta\mc{H}_\mathrm{info}\,,
\end{equation}
which is a generalized version of the maximum work theorem.  If the device begins and ends in equilibrium, $\Delta\mc{F}$ is replaced by the equilibrium free energy difference $\Delta F$.

Equation~\eqref{e34} is similar to a version of the maximum work theorem applicable to systems with external feedback control~\cite{sag10,gra11,vai11,abr12,def12}:
\begin{equation}
\beta \la W\ra\geq \beta \Delta F-\Delta\mc{I} \, .
\end{equation}
Here, $\Delta\mc{I}$ denotes a mutual information that quantifies the quality of the measurements that are performed by an external agent. In Eq.~\eqref{e34}, by contrast, $\Delta\mc{H}_\mathrm{info}$ is the change of Shannon entropy of an explicitly modeled subsystem (our information reservoir), without reference to feedback control. See Ref.~\cite{sag12} for a treatment that combines both perspectives.

\section{Concluding remarks}
\label{sec:conclusions}

By categorizing thermodynamics systems as devices, thermal reservoirs, work reservoirs, and information reservoirs, we have developed an inclusive approach for investigating the thermodynamics of information processing, in which all participating subsystems are explicitly modeled. This approach is based on autonomous evolution under a time-independent Hamiltonian, supplemented by a number of limits, approximations and assumptions, spelled out in Section~\ref{sec:framework}. Our main results in Section~\ref{sec:2ndlaw} generalize the Kelvin-Planck, Clausius, and Carnot statements for cyclic thermodynamic processes, and they support the consensus view~\cite{lan61,penrose_1970,ben82,leff_2003} that the Shannon entropy in a random data set (as encoded by a memory register's information-bearing degrees of freedom, for instance) should be placed on the same footing as the Clausius entropy, when analyzing the second law of thermodynamics.  Thus, for example, work can systematically be extracted from a single heat bath, heat can flow from cold to hot, and the Carnot efficiency can be exceeded, provided these entropy-decreasing consequences are compensated by the writing of information to a memory register.  Section~\ref{sec:maxwork} extends these results to non-cyclic processes, in the form of a generalized maximum work principle.

As mentioned, our derivations have elements in common with previous treatments, particularly those of Refs.~\cite{has10,esp10a,tak10,esp11,sag12}.  However, our focus on a fully autonomous, inclusive framework, on cyclic processes, and on the designation of an information reservoir as a separate element in thermodynamic analyses, distinguishes our approach.  In the spirit of a fully inclusive framework, Maes and Tasaki~\cite{mae07} have derived the maximum work statement of the second law of thermodynamics using a time-independent Hamiltonian.  Their emphasis is on a mathematically rigorous treatment, and does not focus on information-processing.

Very recently, Tasaki \cite{tas13} has analyzed a Hamiltonian model of Maxwell's demon, involving an engine and a memory that interact by the exchange of information, and Barato and Seifert \cite{bar13a} have investigated feedback control with an explicit information reservoir, within the framework of stochastic thermodynamics.


Finally, it is worth mentioning that, at least formally, the present analysis can be extended to quantum-mechanical systems. In place of Hamiltonian dynamics one would use the \textit{unitary} dynamics of the ``universe'' under consideration; the Shannon entropy would be replaced by the von Neumann entropy; and classical ensemble averages would be replaced by quantum expectation values taken with respect to density operators. Aside from these modifications, the mathematical steps in the derivation remain the same. However, because our analysis in this paper has relied heavily on classical reasoning and interpretation, it is not clear whether formally analogous quantal manipulations lead to physically meaningful results.  These conceptual difficulties may perhaps be addressed by appeal to decoherence by an external environment, so as to induce classicality \cite{zurek_2003}, but this is beyond the scope of our approach, which is based on a self-contained, isolated universe. We leave these subtle interpretational issues to future work.

\acknowledgments{SD acknowledges financial support by a fellowship within the postdoc-program of the German Academic Exchange Service (DAAD, contract No D/11/40955).  Both authors acknowledge support from the National Science Foundation (USA) under grant DMR-1206971, and stimulating conversations with Dibyendu Mandal and Jordan M. Horowitz.}

\appendix

\section{Work reservoir}
\label{app:work}

In this appendix we analyze the illustrative example sketched in Fig.~\ref{work_res}. A rarefied gas is confined by a cylinder, whose piston is coupled to a harmonic oscillator.  The Hamiltonian of the work source is
\begin{equation}
\label{a01}
H_\mathrm{work}(X,P)=\frac{P^2}{2 M}+\frac{M \omega^2 X^2}{2}\,,
\end{equation}
where $M$ is the mass of the piston and $\omega$ the angular frequency of the spring. From Hamilton's equations applied to the total Hamiltonian, Eq.~\eqref{e02}, we get, for the piston,
\begin{equation}
\label{a02}
\begin{split}
\dot{X}&=V\hspace{1em} \\
\dot{V}&=-\omega^2 X-\frac{1}{M}\frac{\pd H_0}{\pd X}\,,
\end{split}
\end{equation}
where $V=P/M$ is the piston velocity.  In the limit $M\to\infty$, with the initial position $X_0$ and velocity $V_0$ held fixed, the last term above can be neglected.
The work source then becomes  a work \textit{reservoir} whose time-dependence is given by its free dynamics,
\begin{equation}
\label{a04}
\begin{split}
\lim_{M\rightarrow\infty} X(t)& = X_0 \co{\omega t} -\frac{V_0}{\omega} \si{\omega t}\\
\lim_{M\rightarrow\infty} V(t)& = V_0 \co{\omega t}+X_0\,\omega\si{\omega t}\,,
\end{split}
\end{equation}
which are the periodic solutions used in Section~\ref{subsec:work}.

\section{Information reservoir}
\label{app:info}

Here we provide justification for Eqs.~\eqref{eq:decomposeHinfo} and \eqref{eq:deltaHinfo}, drawing on the assumptions that we have made about the information reservoir.

Given an arbitrary probability distribution $p(\psi)$ on the phase space of the information reservoir, we define marginal and conditional distributions
\begin{equation}
\begin{split}
p_\sigma &= \int {\rm d}\psi \, \delta_{\sigma,\hat\sigma(\psi)} \, p(\psi) \\
p(\psi \vert \sigma) &= \delta_{\sigma,\hat\sigma(\psi)} \, \frac{p(\psi)}{p_\sigma} \, .
\end{split}
\end{equation}
$p_\sigma$ is the probability distribution of informational states; $p(\psi\vert\sigma)$ is the conditional probability distribution of microstates, given an informational state $\sigma$; and
\begin{equation}
p(\psi) = \sum_\sigma p_\sigma \, p(\psi\vert\sigma) \, .
\end{equation}
Introducing the shorthand notation $\int_\sigma {\rm d}\psi \cdots \equiv \int {\rm d}\psi \, \delta_{\sigma,\hat\sigma(\psi)} \cdots$, we obtain
\begin{equation}
\begin{split}
\mc{H}_\mrm{info} &= -\int {\rm d}\psi \, p(\psi) \ln p(\psi) \\
&= -\sum_\sigma p_\sigma \int {\rm d}\psi \, p(\psi\vert\sigma) \ln \left[ \sum_{\sigma^\prime} p_{\sigma^\prime} \, p(\psi\vert\sigma^\prime) \right] \\
&= -\sum_\sigma p_\sigma \int_\sigma {\rm d}\psi \, p(\psi\vert\sigma) \ln \left[ p_\sigma \, p(\psi\vert\sigma) \right] \\
&= -\sum_\sigma p_\sigma \ln p_\sigma \quad -\sum_\sigma p_\sigma \int_\sigma {\rm d}\psi \, p(\psi\vert\sigma) \ln p(\psi\vert\sigma) \\
&= \mc{H}_\mrm{info}^\mrm{IBD} + \sum_\sigma p_\sigma\, \mc{H}_\mrm{info}^\mrm{NBD}(\sigma) \, ,\\
\end{split}
\end{equation}
where $\mc{H}_\mrm{info}^\mrm{NBD}(\sigma)$ is the Shannon entropy associated with the conditional distribution $p(\psi\vert\sigma)$.
Defining $\mc{H}_\mrm{info}^\mrm{NBD} \equiv \sum_\sigma p_\sigma \mc{H}_\mrm{info}^\mrm{NBD}(\sigma)$, we get the decomposition given by Eq.~\eqref{eq:decomposeHinfo}.

Making use of the assumption that the NBD equilibrate rapidly with the thermal reservoir, we replace $\mc{H}_\mrm{info}^\mrm{NBD}(\sigma)$ with its equilibrium value (see Eq.~\eqref{eq:constrainedEnergyEntropy}):
\begin{equation}
\label{eq:Hinfo_appendix}
\mc{H}_\mrm{info} = \mc{H}_\mrm{info}^\mrm{IBD} + \sum_\sigma p_\sigma\, \mc{H}_\mrm{info}^{{\rm eq},\sigma} \, .
\end{equation}
Combining this with the assumption that the equilibrated informational states all have the same entropy ($\mc{H}_\mrm{info}^{{\rm eq},\sigma} = \mc{H}_\mrm{info}^{{\rm eq},\sigma^\prime}$), we conclude that, of the two terms on the right of Eq.~\eqref{eq:Hinfo_appendix}, only the first one changes as the distribution $p_\sigma$ evolves with time.
This finally justifies Eq.~\eqref{eq:deltaHinfo}.


%

\end{document}